\documentclass{article}
\usepackage{spconf,amsmath,graphicx,hyperref}

\usepackage{cite}
\usepackage{amsmath,amssymb,amsfonts}
\usepackage{algorithmic}
\usepackage{graphicx}
\usepackage{textcomp}
\usepackage{xcolor}

\usepackage{booktabs}
\usepackage{booktabs}
\usepackage{threeparttable}
\usepackage{epstopdf}

\usepackage{CJKutf8}
\usepackage{multirow}
\usepackage{epstopdf}
\usepackage{bm}
\usepackage{setspace}
\usepackage{mathrsfs}
\usepackage{times}
\usepackage{latexsym}
\usepackage{stfloats}
\usepackage{cases}
\usepackage{array}
\usepackage{fancyhdr}
\usepackage{subfigure}
\usepackage{balance}
\usepackage{algorithm}
\usepackage{mathtools}
\usepackage{booktabs}
\usepackage{float}
\usepackage{threeparttable}
\usepackage{subfigure}
\usepackage{tikz}
\usepackage{color,xcolor}
\usepackage{amssymb}
\usepackage{footnote}

\title{MPNet: A Robust and Efficient Manifold Pooling Network for Multi-Rhythm EEG Signal Decoding}
\name{Guoqing Cai$^1$   \ \ \ \ \ \ \ \ \ \ 
Kai Zeng$^1$   \ \ \ \ \ \ \ \ \ \  
Shoulin Huang$^3$  \ \ \ \ \ \ \ \ \ \  
Ting Ma$^{1, 2}$ $^{\ast}$ 
\thanks{*Corresponding author}
}
\address{{$^1$Harbin Institute of Technology (Shenzhen), Shenzhen, China}\\
{$^2$Peng Cheng Laboratory, Shenzhen, China}\\
{$^3$Guangxi Normal University, Guilin, China}
}

\begin{document}
\ninept

\maketitle

\begin{abstract}
Deep Riemannian networks provide a powerful framework for Electroencephalography (EEG)  decoding, but their practical applications are severely constrained. 
Accurately decoding EEG signals requires modeling complex temporal dynamics across multiple rhythms, which results in high-dimensional Riemannian inputs and significant computational costs.
To address this, we propose the Manifold Pooling Network (MPNet). 
MPNet uses a rhythm-adaptive convolutional frontend to extract comprehensive time–frequency representations  and generate multi-view Riemannian nodes. 
A novel manifold node pooling layer is then proposed to aggregate these nodes into a single fusion node with a fixed size, 
enabling the following deep Riemannian network to process it with greatly reduced costs.
Experiments on two public EEG datasets show that MPNet achieves state-of-the-art accuracy, 
runs up to 10 times  faster than the comparable Riemannian model, 
and maintains  robust performance under limited-data conditions. 
These findings highlight MPNet's practicality and efficiency for real-world EEG applications.

\end{abstract}


\section{Introduction}
Electroencephalography (EEG), with its non-invasiveness and portability, 
has become an important sensing modality for Brain-Computer Interface (BCI) research, particularly in Motor Imagery (MI) tasks. 
Traditional EEG analysis has often relied on spectral power features, typically derived from power spectral density estimates or wavelet transforms \cite{2004ICASSP1}. 
More recently, deep learning models, ranging from convolutional neural networks (CNN) to Transformers,  
have been developed to directly learn spatio-temporal patterns from raw time series, offering enhanced feature representation capabilities \cite{2018eegnet,2017deepnet,2021fbcnet,2023ifnet}. 
However, such models are highly sensitive to the non-stationarity of EEG signals and lack explicit modeling of inter-channel relationships.

To address these limitations, second-order statistical approaches have been explored as an effective alternative. 
A representative example is the Common Spatial Pattern (CSP) algorithm, which constructs trial-wise covariance matrices to characterize spatial dependencies across channels\cite{2020ICASSP2}. 
Critically, these covariance matrices lie in the space of Symmetric Positive Definite (SPD) matrices, which naturally form a Riemannian manifold \cite{2011multiclass}.
Analyzing SPD matrices in Euclidean space may introduce geometric distortions.  
Consequently, modeling SPD matrices as points on a Riemannian manifold and performing computations within this space better preserves their intrinsic geometric structure, 
leading to improved model robustness and generalization \cite{2024MLCSP}.

Building on this, a series of Riemannian geometry methods have been developed,
employing metrics like the Log-Euclidean or Affine-Invariant distance to classify \cite{2011multiclass, 2024MLCSP, 2021amplitude}. 
While successful in many BCI tasks, they still rely on handcrafted features. 
To this end, SPDNet was proposed as a landmark framework for end-to-end learning on the Riemannian manifold \cite{2017SPDNet}. 
By using manifold-aware layers like BiMap, ReEig, and LogEig, SPDNet allows deep models to process SPD matrices while keeping their geometric structure.
However, the original SPDNet was developed for a single covariance matrix and is unable to model the multi-rhythmic temporal dynamics, which are essential for achieving accurate EEG decoding \cite{2011ICASSP4, 2022eegConformer }.

To address this, several extensions for EEG decoding have been proposed \cite{TensorCSPNet, 2025ICASSP5, 2025EEGSPDNet, 2024ICASSP6, GraphCSPNet}. 
Tensor--CSPNet stacks covariance matrices across frequency bands and time windows, but relies on fixed band definitions and incurs significant computational costs due to high-dimensional SPD inputs\cite{TensorCSPNet}. 
Graph--CSPNet builds graphs over SPD matrices to model time-frequency dependencies, yet remains computationally intensive due to complex manifold operations\cite{GraphCSPNet}. 
EEGSPDNet employs learnable convolutions to generate SPD features with adaptive rhythm optimization. 
Nevertheless, its SPD feature dimensionality increases proportionally with the number of channels, which leads to significant computational costs \cite{2025EEGSPDNet}.
These approaches reveal a critical limitation of deep Riemannian learning for EEG decoding:  prohibitive computational costs prevent their transition from theoretical models to practical applications.

To this end, we propose the Manifold Pooling Network (MPNet) to achieve both accurate EEG decoding and high efficiency in deep Riemannian networks. 
MPNet starts with a rhythm-adaptive convolutional frontend, which uses parallel convolutional branches to capture time-frequency patterns and model cross-frequency interactions. 
Then, we generate a set of multi-view SPD nodes on the Riemannian manifold. 
To overcome the resulting computational bottleneck, we introduce a novel manifold node pooling layer that aggregates these nodes into a single, fixed-size SPD fusion node. 
This key design allows the subsequent deep Riemannian network to process the compact representation efficiently, significantly reducing computational costs. 
As a result, MPNet achieves high classification accuracy and efficiency, making it well-suited for practical BCI applications.

\section{Methodology}
\begin{figure*}
	\centering%
	\includegraphics[scale=0.58,trim=0.1cm 0.6cm 0.4cm 0.6cm,clip]{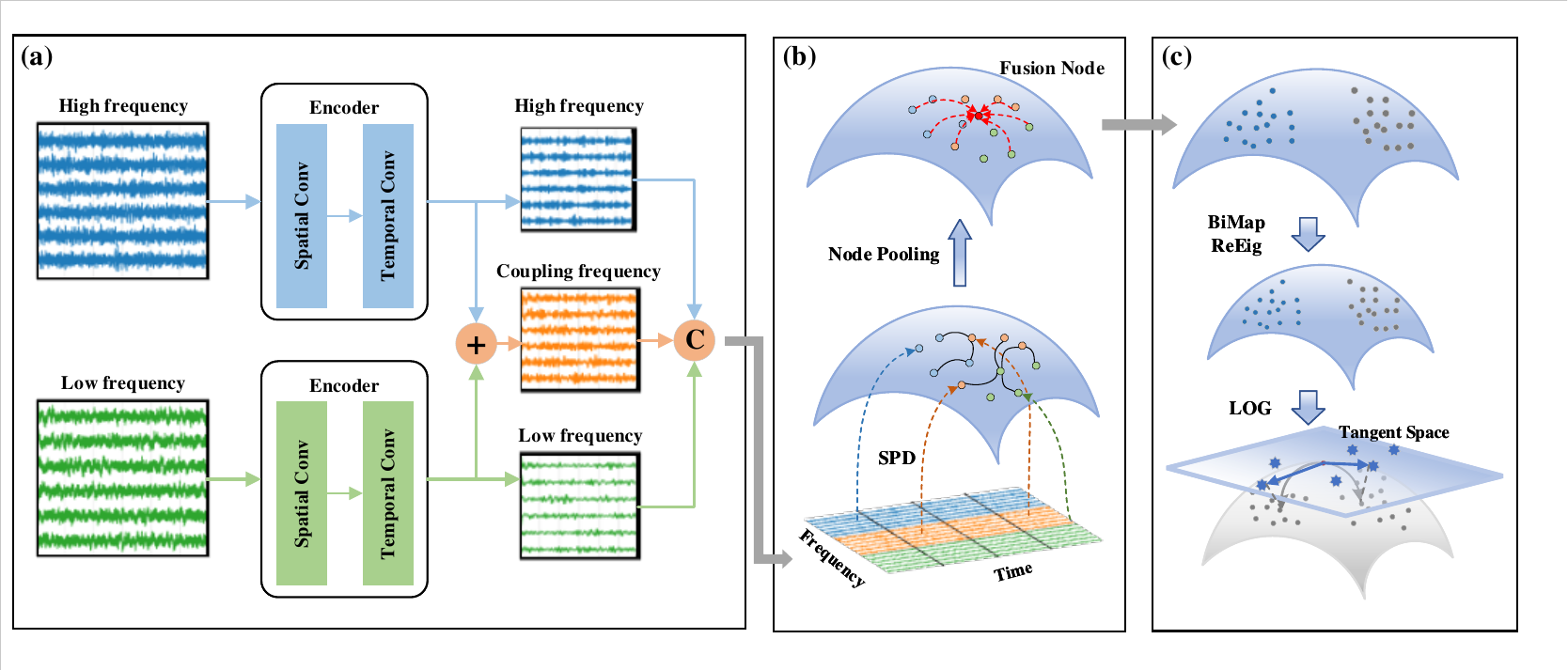}
	\caption{The illustration of the proposed MPNet, which comprises three main stages: 
(a) a \textit{Rhythm-Adaptive Convolutional Module}, 
(b) a \textit{Manifold Node Pooling Module}, and 
(c) a \textit{Riemannian Representation Learning Module}}
    \label{fig:MPNet}
\end{figure*}

The illustration of the proposed MPNet is shown in Fig. \ref{fig:MPNet}.

\subsection{Rhythm-Adaptive Convolutional Module}

The input EEG signal $\mathbf{X} \in \mathbb{R}^{C \times T}$  is first decomposed into two frequency streams using Butterworth filters: 
a low-frequency stream, $\mathbf{X}_{\text{low}}$ ($4$–$17$ Hz) 
and a high-frequency stream $\mathbf{X}_{\text{high}}$ ($13$–$40$ Hz) \cite{2023ifnet}.
$C$ represents the number of channels and  $T$ denotes the number of time samples.
Each rhythm-specific stream is then processed by a dedicated convolutional branch to learn spatio-temporal features.
The branch consists of
a spatial convolution ($C \times 1$ kernel) to capture inter-channel relationships, followed by a temporal convolution ($1 \times 32$ kernel) to perform adaptive rhythm optimization.
Both layers use $D = 60$ output channels, yielding a feature map $\mathbf{H}_k \in \mathbb{R}^{D \times T'}$ for each stream $k$:
\begin{equation}
\mathbf{H}_k = \text{Conv}_{\text{time}} \left( \text{Conv}_{\text{spatial}}(\mathbf{X}_k) \right), \quad k \in \{ \text{low}, \text{high} \}
\end{equation}
To enhance rhythm interactions, we compute a coupling branch by summing the two filtered signals \cite{2023ifnet}:
\begin{equation}
\mathbf{H}_{\text{coup}} = \mathbf{H}_{\text{low}} + \mathbf{H}_{\text{high}}  \in \mathbb{R}^{D \times T'}
\end{equation}
The final output of this stage is a set of three distinct feature streams, each providing a unique view of the brain's rhythmic activity:
\begin{equation}
\{ \mathbf{H}_{\text{low}}, \mathbf{H}_{\text{high}}, \mathbf{H}_{\text{coup}} \}
\end{equation}

\subsection{Manifold Node Pooling Module}

This core module aggregates the multi-view SPD nodes into a single SPD matrix with a fixed size, enabling efficient downstream Riemannian processing.
Each rhythmic stream $\mathbf{H}_k$ is divided into $W=4$ non-overlapping time windows following the setting in \cite{GraphCSPNet}. 
For each windowed segment $\mathbf{H}_{k,w} \in \mathbb{R}^{D \times T_w}$, we compute its covariance matrix: 
\begin{equation}
\mathbf{S}_i = \frac{1}{T_w - 1} \mathbf{H}_{k,w} \mathbf{H}_{k,w}^\top \in \mathbb{R}^{D \times D}
\end{equation}
where $T_w$ is the number of time samples in the local time windows, $ k \in \{ \text{low}, \text{high}, \text{coup} \}$ and $ w=1,...,W$.
This results in a total of $N = 3 \times W = 12$ SPD matrices: $\{\mathbf{S}_1, \mathbf{S}_2, \dots, \mathbf{S}_N\}$.
Next, we propose a manifold node pooling layer that aggregates these matrices into a fusion node $\mathbf{S}_{\text{pooled}}$. 
We consider four pooling strategies:

{(1)} Euclidean Mean Pooling (EM):
\begin{equation}
\textstyle  \mathbf{S}_{\text{pooled}} =  \frac{1}{N}\sum_{i=1}^{N} \mathbf{S}_i
\end{equation}

{(2)} Weighted Euclidean Mean Pooling (WEM):
\begin{equation}
\textstyle  \mathbf{S}_{\text{pooled}} = \sum_{i=1}^{N} \alpha_i \mathbf{S}_i, \quad    \boldsymbol{\alpha} = \text{softmax}(\mathbf{w})
\end{equation}

 {(3)} Approximate Riemannian Mean Pooling (RM) \cite{2025RM}:
\begin{equation}
\textstyle  \mathbf{S}_{\text{pooled}} = \exp\left( \frac{1}{N} \sum_{i=1}^{N} \log(\mathbf{S}_i) \right)
\end{equation}

{(4)} Weighted Riemannian Mean Pooling (WRM):
\begin{equation}
\textstyle  \mathbf{S}_{\text{pooled}} = \exp\left( \sum_{i=1}^{N} \alpha_i \log(\mathbf{S}_i) \right), \quad  \boldsymbol{\alpha} = \text{softmax}(\mathbf{w})
\end{equation}
where $\mathbf{w} \in \mathbb{R}^{N}$ is a learnable vector.  

This stage ensures that the next module receives a compact and  fixed-size representation $\mathbf{S}_{\text{pooled}} \in \text{SPD}(D)$, regardless of the number of rhythms or windows used.

\subsection{Riemannian Representation Learning Module}

This module extracts discriminative, high-level features from the global SPD representation $\mathbf{S}_{\text{pooled}}$ using a geometry-preserving deep Riemannian network, SPDNet \cite{2017SPDNet}. 
SPDNet operates entirely on the Riemannian manifold and is composed of a series of manifold-aware layers:

\textbf{BiMap Layer:} A bilinear mapping that projects the input SPD matrix $\mathbf{S}_{\text{in}}  \in \mathbb{R}^{d \times d}$ to a lower-dimensional SPD space:
\begin{equation}
\mathbf{S}_{\text{out}} = \mathbf{W} \mathbf{S}_{\text{in}} \mathbf{W}^\top \in \mathbb{R}^{d' \times d'}
\end{equation}
where $\mathbf{W} \in \mathbb{R}^{d' \times d}$ is a learnable weight matrix.
${\mathbf{S}}_{\text{in}}$ and ${\mathbf{S}}_{\text{out}}$ correspond to the input and output of the BiMap layer.

\textbf{ReEig Layer:} An eigenvalue rectification function to ensure numerical stability:
\begin{equation}
\mathbf{S}_{\text{out}} = \mathbf{U} \, \text{diag}(\max(\boldsymbol{\lambda}, \epsilon)) \, \mathbf{U}^\top
\end{equation}
where $\mathbf{S}_{\text{in}} = \mathbf{U} \text{diag}(\boldsymbol{\lambda}) \mathbf{U}^\top$ and $\epsilon > 0$ is a small threshold. 
Here, ${\mathbf{S}}_{\text{in}}$ and ${\mathbf{S}}_{\text{out}}$ correspond to the input and output of the ReEig layer.

\textbf{LogEig Layer:} Maps the SPD matrix to the tangent  space (Euclidean space) by computing the matrix logarithm:
\begin{equation}
\mathbf{X}_{\text{log}} = \log(\mathbf{S}_{\text{in}}) = \mathbf{U} \, \text{diag}(\log(\boldsymbol{\lambda})) \, \mathbf{U}^\top
\end{equation}

The result $\mathbf{X}_{\text{log}} \in \mathbb{R}^{d \times d}$ is then vectorized:
\begin{equation}
\mathbf{x} = \text{vec}(\mathbf{X}_{\text{log}}) \in \mathbb{R}^{\frac{d \times {(d+1)}}{2}}
\end{equation}

In practice, multiple BiMap–ReEig blocks can be stacked to compress SPD features while preserving geometric consistency.
In the study, we use two such blocks.
The first block maps from $D=60$ to $D'=30$.
The second maps from $D'=30$ to $D''=15$.
Then, a LogEig layer is applied to project the final SPD matrix onto the tangent space. 
The output is passed through a fully connected layer for classification.

\section{Experiment}

\subsection{Datasets and Preprocessing}

To evaluate the effectiveness of the proposed MPNet, we conducted experiments on two publicly available MI-EEG datasets: \textbf{BCI Competition IV 2a (BCI-IV 2a)} \cite{2012BCI} and \textbf{OpenBMI}\cite{2019OpenBMI}. 
BCI-IV 2a contains EEG recordings from 9 healthy subjects performing four MI tasks, collected using 22 electrodes at a sampling rate of 250 Hz. 
Each subject completed two sessions on different days.
OpenBMI includes data from 54 subjects performing two MI tasks (left vs. right hand), recorded using a 62-channel EEG cap at 1000 Hz, also over two sessions.

We adopted a consistent preprocessing pipeline. For each trial, EEG segments from 0 to 4 seconds after cue onset were extracted.
 Signals from OpenBMI  were downsampled to 250 Hz to match BCI-IV 2a.  
 A 5th-order Butterworth bandpass filter (4–40 Hz) was then applied.
We used the first session for training and the second for testing.

\subsection{Compared Methods}
We compared MPNet against several state-of-the-art models spanning both convolutional and Riemannian geometry-based approaches.
CNN-based models include EEGNet \cite{2018eegnet}, 
ShallowNet \cite{2017deepnet}, FBCNet \cite{2021fbcnet}, IFNet \cite{2023ifnet}, and EEGConformer \cite{2022eegConformer}, 
which range from lightweight convolutional designs to hybrid CNN-Transformer models. 
 For Riemannian models, we included 
 GREEN \cite{2025green}  and EEGSPDNet \cite{2025EEGSPDNet}, which integrate the convolutional frontend with SPDNet  for end-to-end manifold learning.  
 These models represent diverse and competitive design paradigms for MI decoding, providing a comprehensive analysis for evaluating MPNet.

\subsection{Implementation and Evaluation Metrics}

In this study, the cross-entropy loss was used across all models. 
We adopted the Adam optimizer with a learning rate of 0.001 for all CNN-based models. For Riemannian models and the SPDNet component of MPNet, 
we used the Riemannian Adam optimizer as recommended in \cite{2025EEGSPDNet}, also with a learning rate of 0.001.
Models were trained for up to 1500 epochs, with early stopping based on the loss using a patience of 150 epochs. 
We employed accuracy (ACC), Cohen’s kappa, and F1-score (F1) to evaluate performance.

\section{Results}

\subsection{Analysis of Manifold Pooling Strategies}

\begin{table}[htbp]
\footnotesize
\centering
\renewcommand\arraystretch{1} 
\tabcolsep=0.08cm 
    \caption{Ablation Study of Manifold Node Pooling Strategies on the BCI-IV 2a and OpenBMI Dataset. 
             The best result is highlighted in bold, and the second-best is underlined.}
    \label{tab:pooling_ablation}
    \begin{tabular}{l|ccc|ccc}
      \toprule
      & \multicolumn{3}{c|}{\textbf{BCI-IV 2a Dataset}} & \multicolumn{3}{c}{\textbf{OpenBMI Dataset}} \\
      \cmidrule(lr){2-4} \cmidrule(lr){5-7}
      Methods & ACC (\%) & Kappa (\%) & F1 (\%) & ACC (\%) & Kappa (\%) & F1 (\%) \\
      \hline
     
      EM                & \underline{80.71} & \underline{74.28} & \underline{80.37}   & \textbf{79.60} & \textbf{59.20} & \textbf{79.33}   \\
      WEM             & 80.40 & 73.86  & 80.10  & \underline{79.10} & \underline{58.20}  & \underline{78.78}    \\
      RM              & 80.03 & 73.37 & 79.77   & 78.77 & 57.54 & 78.51\\
      WRM             & \textbf{80.97} & \textbf{74.62} & \textbf{80.74}   & 78.96 & 57.92 & 78.47 \\ 
       w/o MP		  & 76.35 &  68.47 & 74.68   & 76.05 &  51.20  &  75.81   \\
      \bottomrule
    \end{tabular}
\end{table}

We investigate the impact of manifold node pooling (MP) and compare different pooling operators in MPNet.
Table~\ref{tab:pooling_ablation} reports the results on BCI-IV~2a and OpenBMI.
We first compare four manifold pooling operators in MPNet, including EM, WEM, RM, and WRM, as reported in Table~\ref{tab:pooling_ablation}.
On BCI-IV~2a, all operators yield competitive performance, and WRM achieves the highest accuracy of 80.97\%.
On OpenBMI, EM performs best and reaches 79.60\% in accuracy, 59.20\% in Kappa, and 79.33\% in F1.
Given the small margin among operators on BCI-IV~2a and the clear advantage of EM on OpenBMI, we choose EM as the default pooling strategy for MPNet.

We then evaluate the contribution of manifold pooling by removing the pooling module.
Without pooling (w/o MP), performance consistently degrades on both datasets.
On BCI-IV~2a, accuracy drops from 80.71\% to 76.35\%, and the corresponding Kappa and F1 decrease from 74.28\% to 68.47\% and from 80.37\% to 74.68\%.
On OpenBMI, accuracy decreases from 79.60\% to 76.05\%, and Kappa and F1 decrease from 59.20\% to 51.20\% and from 79.33\% to 75.81\%.
These results indicate that pooling is essential for consolidating time--rhythm SPD nodes into a unified and robust representation.
Besides, we analyze efficiency under the default EM setting.
On BCI-IV~2a, removing pooling increases the training time from 0.17~s/epoch to 0.62~s/epoch and the testing time from 0.07~s/epoch to 0.25~s/epoch.
On OpenBMI, the training time increases from 0.28~s/epoch to 0.93~s/epoch.
Overall, manifold pooling improves both recognition performance and computational efficiency, and EM provides the best balance across datasets.

\subsection{Classification Performance}

\begin{table*}[htbp]
\footnotesize
\centering
\renewcommand\arraystretch{1}
\tabcolsep=0.14cm 
\caption{Comprehensive performance comparison (Mean $\pm$ Std) on the BCI-IV 2a and OpenBMI datasets. 
             The best result is highlighted in bold, and the second-best is underlined.}
\label{tab:results_combined}
\begin{tabular}{l|cccl|cccl}
    \toprule
    & \multicolumn{4}{c|}{\textbf{BCI-IV 2a Dataset}} & \multicolumn{4}{c}{\textbf{OpenBMI Dataset}} \\
    \cmidrule(lr){2-5} \cmidrule(lr){6-9}
    Methods & ACC (\%) & Kappa (\%) & F1 (\%) & $p$-value & ACC (\%) & Kappa (\%) & F1 (\%) & $p$-value \\
    \midrule
    EEGNet (2018)        & 70.52 $\pm$ 15.25 & 60.70 $\pm$ 20.33 & 70.11 $\pm$ 15.51 & 0.0039$^{**}$               & 70.75 $\pm$ 14.44  & 41.50 $\pm$ 19.25 & 70.37 $\pm$ 14.75 & $<0.0001$$^{**}$ \\
    
    ShallowNet (2017)   & 72.34 $\pm$ 11.96 & 63.12 $\pm$ 15.95 & 71.84 $\pm$ 12.31 & 0.0020$^{**}$               & 74.99 $\pm$ 13.13   & 49.98 $\pm$ 17.51 & 74.64 $\pm$ 13.49 & $<0.0001$$^{**}$ \\
    
    FBCNet (2021)       & 77.53 $\pm$ 11.82 & 70.04 $\pm$ 15.76 & 77.01 $\pm$ 12.28 & 0.0098$^{**}$                & 74.04 $\pm$ 13.98  & 48.08 $\pm$ 18.64 & 73.66 $\pm$ 14.33 & $<0.0001$$^{**}$ \\

    IFNet (2023)        & 78.24 $\pm$ 10.63 & 70.98 $\pm$ 14.17 & 78.01 $\pm$ 10.76 & 0.0273$^{*}$                & 78.13 $\pm$ 13.50   & 56.26 $\pm$ 18.00 & 77.95 $\pm$ 13.89 & 0.0120$^{*}$ \\

    EEGConformer (2023) & 79.09 $\pm$ 11.48 & 72.12 $\pm$ 15.31 & 78.70 $\pm$ 11.97 & 0.0371$^{*}$                & \underline{78.57 $\pm$ 13.41}   & \underline{57.14 $\pm$ 17.88} & \underline{78.31 $\pm$ 13.67} & 0.0205$^{*}$ \\
        
    GREEN (2025)        & 64.83 $\pm$ 13.95 & 53.11 $\pm$ 18.60 & 64.32 $\pm$ 14.33 & 0.0020$^{**}$               & 70.69 $\pm$ 13.91   & 41.38 $\pm$ 18.55 & 69.87 $\pm$ 14.93 & $<0.0001$$^{**}$ \\
    
    EEGSPDNet (2025)   & \underline{79.22 $\pm$ 10.99} & \underline{72.30 $\pm$ 14.65} & \underline{79.02 $\pm$ 11.22} & 0.0645         & 76.48 $\pm$ 13.47 & 52.96 $\pm$ 17.97 & 75.74 $\pm$ 14.36 & $<0.0001$$^{**}$ \\

    MPNet (Our) & \textbf{80.71 $\pm$ 11.35} & \textbf{74.28 $\pm$ 15.13} & \textbf{80.37 $\pm$ 11.78} & -- & \textbf{79.60 $\pm$ 13.18} & \textbf{59.20 $\pm$ 17.57} & \textbf{79.33 $\pm$ 13.49} & -- \\
    \bottomrule
    
\end{tabular}
\end{table*}

Secondly, we conducted a comprehensive classification performance comparison, with the results summarized in Table \ref{tab:results_combined}, 
where the statistical significance of the results was evaluated using the Wilcoxon signed-rank test ($^{**}$ indicates $p < 0.01$, $^{*}$ indicates $p < 0.05$).
The results  clearly demonstrate the superiority of our proposed MPNet. 
On both the BCI-IV 2a and OpenBMI datasets, MPNet consistently achieves the highest performance across all metrics, obtaining mean accuracies of 80.71\% and 79.60\%, respectively. 
It is noteworthy that on the BCI-IV 2a dataset, the Riemannian-based method, EEGSPDNet, secures the second-best performance, confirming the inherent advantage of modeling data on the Riemannian manifold. 
Moreover, our proposed MPNet surpasses it and all other competing methods, with statistically significant gains ($p < 0.05$) over most models.

\subsection{Robustness Analysis}

\begin{figure}
	\centering%
	\includegraphics[scale=0.54,trim=0cm 0.3cm 0.1cm 0.2cm,clip]{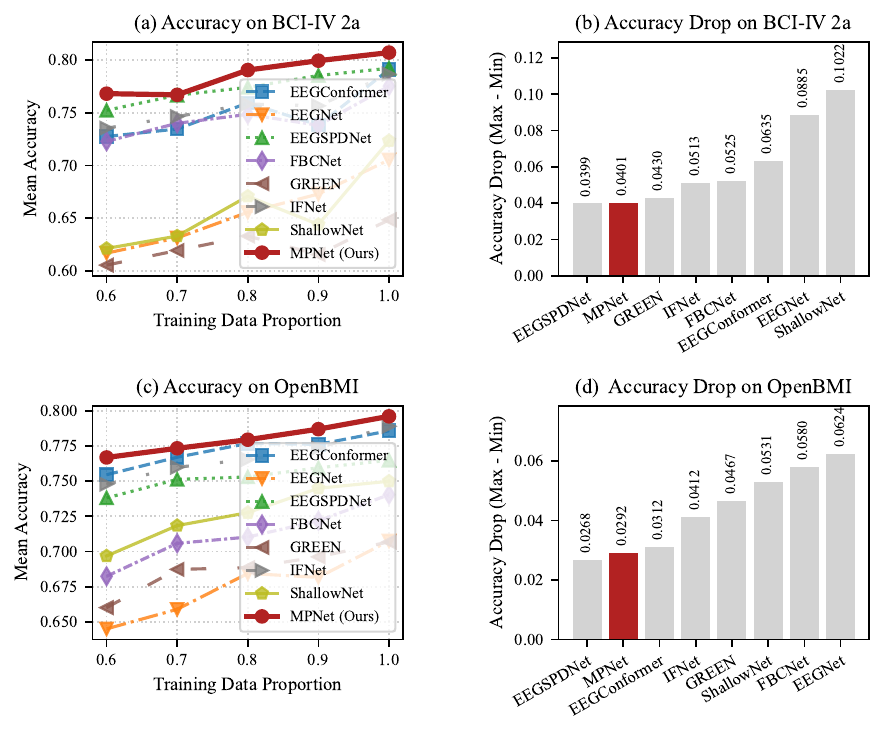}
	\caption{Accuracy and stability across varying training data proportions on the two datasets.}
    \label{fig:Robustness}
\end{figure}

We further analyze the robustness of MPNet by adjusting the training data ratio from 60\% to 100\% on BCI-IV 2a and OpenBMI, as shown in Fig. \ref{fig:Robustness}.
MPNet consistently maintains the highest accuracy across all training data ratios, with a particularly large margin under limited-data conditions (e.g., 60\%).
Stability analysis further indicates that MPNet is highly stable and achieves the second-smallest performance drop. 
Notably, EEGSPDNet, which also adopts a Riemannian approach, shows the lowest drop.
In contrast, Euclidean models such as EEGNet and ShallowNet experience the largest performance drop, exceeding 8\% in BCI-IV 2a and 5\% in OpenBMI.
These results highlight the robustness advantage of Riemannian methods, as both EEGSPDNet and MPNet maintain high performance even under limited training data.

\subsection{Computational Efficiency Analysis}

\begin{figure}
	\centering%
	\includegraphics[scale=0.54,trim=0cm 0.3cm 0cm 0.15cm,clip]{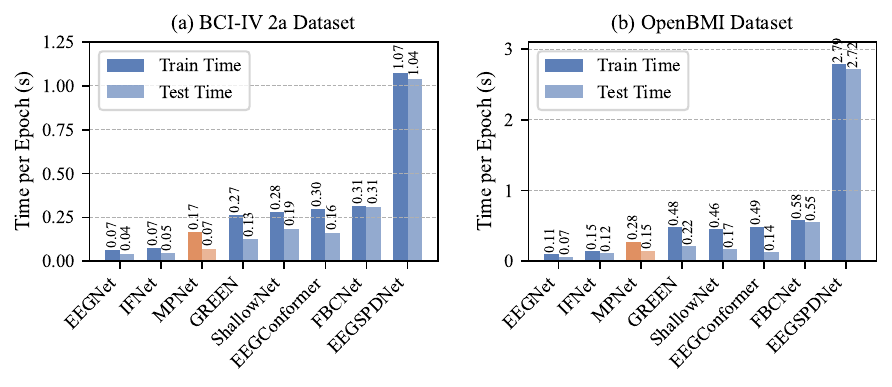}
	\caption{Computational cost comparison. All models were evaluated on an NVIDIA RTX 2080 Ti GPU with 12 GB of memory.}
    \label{fig:Efficiency}
\end{figure}

Beyond accuracy, computational efficiency is critical for practical EEG-based BCI systems.
As shown in Fig.~\ref{fig:Efficiency}, MPNet achieves high training and inference efficiency.
Its time cost is only slightly higher than lightweight CNN baselines such as EEGNet and IFNet, while delivering substantially better accuracy and robustness as reported in Table~\ref{tab:results_combined} and Fig.~\ref{fig:Robustness}.
In contrast, EEGSPDNet incurs the highest computational cost. Its high-dimensional Riemannian inputs lead to nearly one order of magnitude longer running time than MPNet.
This efficiency is enabled by our manifold node pooling module, 
which compresses high-dimensional SPD nodes into a compact, fixed-size fusion node while preserving discriminative power.

We further report theoretical complexity to complement the runtime evaluation.
On BCI-IV~2a, MPNet requires 0.95M FLOPs and 8.24K parameters, while IFNet requires 9.86M FLOPs and 9.34K parameters.
MPNet therefore has comparable model size but substantially lower FLOPs.
Besides, MPNet achieves higher accuracy than IFNet and shows stronger robustness under moderately reduced training data.

\section{Conclusion}

In this study, we proposed MPNet, a novel convolutional Riemannian network designed to address the computational limitations of existing Riemannian models in MI-EEG decoding.
MPNet integrates a rhythm-adaptive frontend with a novel manifold node pooling module, 
which transforms a set of time-frequency SPD nodes into a fixed-size SPD representation.
It enables the following  deep Riemannian learning with significantly reduced computational costs.
Experimental results confirm that MPNet achieves state-of-the-art accuracy with significantly improved computational efficiency, 
while also demonstrating superior robustness, especially in limited-data  scenarios. 
These findings confirm that the proposed MPNet architecture is a promising solution for practical EEG-based BCI systems, 
where decoding accuracy, computational efficiency, and robustness are simultaneously critical requirements.

\bibliographystyle{IEEEtran}%
\bibliography{refs}

@INPROCEEDINGS{2004ICASSP1,
  author={Fatourechi, M. and Mason, S.G. and Birch, G.E. and Ward, R.K.},
  booktitle={2004 IEEE International Conference on Acoustics, Speech, and Signal Processing}, 
  title={A wavelet-based approach for the extraction of event related potentials from EEG}, 
  year={2004},
  volume={2},
  number={},
  pages={ii-737},
  doi={10.1109/ICASSP.2004.1326363}}

@article{2018eegnet,
  title={EEGNet: a compact convolutional neural network for EEG-based brain--computer interfaces},
  author={Lawhern, Vernon J and Solon, Amelia J and Waytowich, Nicholas R and Gordon, Stephen M and Hung, Chou P and Lance, Brent J},
  journal={Journal of neural engineering},
  volume={15},
  number={5},
  pages={056013},
  year={2018},
  publisher={iOP Publishing}
}

@article{2017deepnet,
  title={Deep learning with convolutional neural networks for EEG decoding and visualization},
  author={Schirrmeister, Robin Tibor and Springenberg, Jost Tobias and Fiederer, Lukas Dominique Josef and Glasstetter, Martin and Eggensperger, Katharina and Tangermann, Michael and Hutter, Frank and Burgard, Wolfram and Ball, Tonio},
  journal={Human brain mapping},
  volume={38},
  number={11},
  pages={5391--5420},
  year={2017},
  publisher={Wiley Online Library}
}

@article{2021fbcnet,
  title={FBCNet: A multi-view convolutional neural network for brain-computer interface},
  author={Mane, Ravikiran and Chew, Effie and Chua, Karen and Ang, Kai Keng and Robinson, Neethu and Vinod, A Prasad and Lee, Seong-Whan and Guan, Cuntai},
  journal={arXiv preprint arXiv:2104.01233},
  year={2021}
}

@article{2023ifnet,
  title={IFNet: An interactive frequency convolutional neural network for enhancing motor imagery decoding from EEG},
  author={Wang, Jiaheng and Yao, Lin and Wang, Yueming},
  journal={IEEE Transactions on Neural Systems and Rehabilitation Engineering},
  volume={31},
  pages={1900--1911},
  year={2023},
  publisher={IEEE}
}

@article{2022eegConformer,
  title={EEG conformer: Convolutional transformer for EEG decoding and visualization},
  author={Song, Yonghao and Zheng, Qingqing and Liu, Bingchuan and Gao, Xiaorong},
  journal={IEEE Transactions on Neural Systems and Rehabilitation Engineering},
  volume={31},
  pages={710--719},
  year={2022},
  publisher={IEEE}
}

@INPROCEEDINGS{2020ICASSP2,
  author={Jiang, Aimin and Shang, Jing and Cheng, Weigao and Liu, Xiaofeng and Kwan, Hon Keung and Zhu, Yanping},
  booktitle={2020 IEEE International Conference on Acoustics, Speech and Signal Processing (ICASSP)}, 
  title={Sparse CSP Algorithm via Joint Spatio-Temporal Filtering}, 
  year={2020},
  volume={},
  number={},
  pages={1035-1039},
  doi={10.1109/ICASSP40776.2020.9054526}}

@INPROCEEDINGS{2024ICASSP6,
  author={Qin, Ruihan and Song, Zhenxi and Ren, Huixia and Pei, Zian and Zhu, Lin and Shi, Xue and Guo, Yi and Liu, Honghai and Zhang, Min and Zhang, Zhiguo},
  booktitle={2024 IEEE International Conference on Acoustics, Speech and Signal Processing (ICASSP)}, 
  title={BNMTrans: A Brain Network Sequence-Driven Manifold-Based Transformer for Cognitive Impairment Detection Using EEG}, 
  year={2024},
  volume={},
  number={},
  pages={2016-2020},
  doi={10.1109/ICASSP48485.2024.10447106}}

@article{2025green,
  title={GREEN: A lightweight architecture using learnable wavelets and Riemannian geometry for biomarker exploration with EEG signals},
  author={Paillard, Joseph and Hipp, J{\"o}rg F and Engemann, Denis A},
  journal={Patterns},
  volume={6},
  number={3},
  year={2025},
  publisher={Elsevier}
}

@article{2025EEGSPDNet,
  title={Deep riemannian networks for end-to-end eeg decoding},
  author={Wilson, Daniel and Schirrmeister, Robin T and Gemein, Lukas AW and Ball, Tonio},
  journal={Imaging Neuroscience},
  volume={3},
  pages={imag\_a\_00511},
  year={2025},
  publisher={MIT Press 255 Main Street, 9th Floor, Cambridge, Massachusetts 02142, USA~…}
}

@INPROCEEDINGS{2011ICASSP4,
  author={Higashi, Hiroshi and Tanaka, Toshihisa},
  booktitle={2011 IEEE International Conference on Acoustics, Speech and Signal Processing (ICASSP)}, 
  title={Classification by weighting for spatio-frequency components of EEG signal during motor imagery}, 
  year={2011},
  volume={},
  number={},
  pages={585-588},
  doi={10.1109/ICASSP.2011.5946471}}

@INPROCEEDINGS{2025ICASSP5,
  author={Wang, Zirui and Song, Zhenxi and Guo, Yi and Liu, Yuxin and Xu, Guoyang and Zhang, Min and Zhang, Zhiguo},
  booktitle={2025 IEEE International Conference on Acoustics, Speech and Signal Processing (ICASSP)}, 
  title={EEG-ReMinD: Enhancing Neurodegenerative EEG Decoding through Self-Supervised State Reconstruction-Primed Riemannian Dynamics}, 
  year={2025},
  volume={},
  number={},
  pages={1-5},
  doi={10.1109/ICASSP49660.2025.10887579}}

@inproceedings{2017SPDNet,
  title={A riemannian network for spd matrix learning},
  author={Huang, Zhiwu and Van Gool, Luc},
  booktitle={Proceedings of the AAAI conference on artificial intelligence},
  volume={31},
  number={1},
  year={2017}
}

@article{2024MLCSP,
  title={Manifold learning-based common spatial pattern for EEG signal classification},
  author={Cai, Guoqing and Zhang, Fenghui and Yang, Bolun and Huang, Shoulin and Ma, Ting},
  journal={IEEE Journal of Biomedical and Health Informatics},
  volume={28},
  number={4},
  pages={1971--1981},
  year={2024},
  publisher={IEEE}
}

@article{2021amplitude,
  title={Amplitude-phase information measurement on Riemannian manifold for motor imagery-based BCI},
  author={Huang, Shoulin and Cai, Guoqing and Wang, Tong and Ma, Ting},
  journal={IEEE Signal Processing Letters},
  volume={28},
  pages={1310--1314},
  year={2021},
  publisher={IEEE}
}

@article{2011multiclass,
  title={Multiclass brain--computer interface classification by Riemannian geometry},
  author={Barachant, Alexandre and Bonnet, St{\'e}phane and Congedo, Marco and Jutten, Christian},
  journal={IEEE Transactions on Biomedical Engineering},
  volume={59},
  number={4},
  pages={920--928},
  year={2011},
  publisher={Ieee}
}

@ARTICLE{TensorCSPNet,
  author={Ju, Ce and Guan, Cuntai},
  journal={IEEE Transactions on Neural Networks and Learning Systems}, 
  title={Tensor-CSPNet: A Novel Geometric Deep Learning Framework for Motor Imagery Classification}, 
  year={2023},
  volume={34},
  number={12},
  pages={10955-10969},
  doi={10.1109/TNNLS.2022.3172108}}

@ARTICLE{GraphCSPNet,
  author={Ju, Ce and Guan, Cuntai},
  journal={IEEE Transactions on Neural Networks and Learning Systems}, 
  title={Graph Neural Networks on SPD Manifolds for Motor Imagery Classification: A Perspective From the Time Frequency Analysis}, 
  year={2024},
  volume={35},
  number={12},
  pages={17701-17715},
  doi={10.1109/TNNLS.2023.3307470}}

@ARTICLE{2025RM,
  author={Shi, Yuxuan and Jiang, Aimin and Zhong, Ju and Li, Min and Zhu, Yanping},
  journal={IEEE Journal of Biomedical and Health Informatics}, 
  title={Multiclass Classification Framework of Motor Imagery EEG by Riemannian Geometry Networks}, 
  year={2025},
  volume={29},
  number={2},
  pages={935-947},
  doi={10.1109/JBHI.2024.3496757}}

@article{2019OpenBMI,
  title={EEG dataset and OpenBMI toolbox for three BCI paradigms: an investigation into BCI illiteracy},
  author={ Min-Ho, Lee  and  O-Yeon, Kwon  and  Yong-Jeong, Kim  and  Hong-Kyung, Kim  and  Young-Eun, Lee  and  John, Williamson  and  Siamac, Fazli  and  Seong-Whan, Lee },
  journal={Gigaence},
  number={5},
  pages={5},
  year={2019},
}

@ARTICLE{2012BCI,
AUTHOR={Tangermann, Michael  and Müller, Klaus-Robert  and Aertsen, Ad  and Birbaumer, Niels  and Braun, Christoph  and Brunner, Clemens  and Leeb, Robert  and Mehring, Carsten  and Miller, Kai J. and Mueller-Putz, Gernot  and Nolte, Guido  and Pfurtscheller, Gert  and Preissl, Hubert  and Schalk, Gerwin  and Schlögl, Alois  and Vidaurre, Carmen  and Waldert, Stephan  and Blankertz, Benjamin },
TITLE={Review of the BCI Competition IV},
JOURNAL={Frontiers in Neuroscience},
VOLUME={Volume 6 - 2012},
YEAR={2012},
DOI={10.3389/fnins.2012.00055},
ISSN={1662-453X},
}

\end{document}